\documentclass[aps,pra,twocolumn,showpacs,superscriptaddress]{revtex4}
\usepackage{graphicx}
\usepackage{epstopdf}
\usepackage{amsmath}
\usepackage{amssymb}
\usepackage{float}

\input{epsf}
\begin{document}

\title{Tunable high-temperature thermodynamics of 
weakly-interacting
dipolar gases}

\author{K.~M. Daily}
\affiliation{Department of Physics,
Purdue University,
West Lafayette, Indiana 47907, USA}
\affiliation{Department of Physics and Astronomy,
Washington State University,
  Pullman, Washington 99164-2814, USA}
\author{D. Blume}
\affiliation{Department of Physics and Astronomy,
Washington State University,
  Pullman, Washington 99164-2814, USA}

\date{\today}

\begin{abstract}
We consider dilute gases of dipolar bosons or fermions
in the high-temperature limit
in a spherically symmetric harmonic trapping potential.
We examine the system using a virial expansion up to second order
in the fugacity.
Using the Born approximation and assuming purely dipolar interactions, 
we find that the second-order virial coefficient 
for both bosons and fermions depends quadratically 
on the dipole length and is negative at high temperature,
indicating that to lowest order 
in the dipole-dipole interactions
the dipolar single-component quantum gases are repulsive.
If the $s$-wave scattering length for the bosonic system is
tunable and its absolute value is made small,
then the $s$-wave interactions dominate and 
the dipolar gas behaves like a weakly-interacting Bose gas
with  isotropic $s$-wave interactions. 
If the generalized scattering lengths 
for the fermionic system are tunable,
then the dipole length can enter linearly 
in the virial equation of state,
enhancing the dipole-dipole effects in the thermodynamic observables.
\end{abstract}


\maketitle

\section{Introduction}
There has been much interest in the past few years 
in ultracold systems of 
dipoles~\cite{bara08,LahayeRPP2009,UlmanisChemRev2012}.
Anisotropic long-range interactions can arise due to the 
interactions between the magnetic moments of atoms
or due to the interactions between 
heteronuclear diatomic molecules (in this case, each molecule is
treated as an electric dipole).
Experimentally, dipole-dipole interactions play a prominent
role in the dynamics of ultracold atomic clouds consisting of
Cr~\cite{LahayeNature2007,LahayePRL2008,Beaufils2008R}, 
Dy~\cite{LuLev2011,LuLev2012,SchmittPfau2013}, and 
Er~\cite{Aikawa2012}. 
In these systems,
the effects of the dipole-dipole interactions 
have been observed in the degenerate regime
via expansion cloud 
imaging~\cite{StuhlerPRL1995,LahayePRL2008,LuLev2011,Aikawa2012}.
Additionally, polar molecules such as
KRb~\cite{NeyenhuisPRL2012},
RbCs~\cite{SageDeMillePRL2005} and
LiCs~\cite{DeiglmayrPRL2008}
have been created and the effects of the anisotropic interactions 
have been observed in KRb clouds with phase space densities
just above degeneracy~\cite{ni10,miranda}.

It remains an open question whether the effects from 
the anisotropic and long-range dipole-dipole interactions
can be observed in thermodynamic observables
at high temperature, i.e., at temperatures
above the degeneracy temperature.
For isotropic short-range interactions,
the high-temperature behavior of two-component
Fermi gases is well described by the virial equation of 
state~\cite{LiuPRL2009,HuNJP2010}.
At unitarity, i.e., for infinitely large
interspecies $s$-wave scattering length, the
virial equation of state has been validated through careful 
measurements~\cite{NascimbeneNature2010,KuScience2012}.
This paper derives the virial equation of state
for dipolar gases and investigates the interplay between
the short-range and long-range interactions. 
Specifically, we focus on the temperature regime above
quantum degeneracy and show that signatures
of the dipole-dipole interactions can be enhanced 
if one of the generalized 
scattering lengths in a low partial wave channel is tuned.

We consider a single-component 
gas of aligned dipolar bosons or fermions under spherically symmetric
harmonic confinement
and calculate the second-order virial coefficient
for weak dipole-dipole interactions.
If the dipole-dipole scattering properties are calculated in the
first-order Born approximation
and if the dipole length $D$ is small
compared to the harmonic oscillator length
$a_{\rm ho}$,
then the second-order virial coefficient $\Delta b_2$, 
which accounts for the two-body interactions, is negative 
at large $T$ for
both identical bosons and identical fermions
and depends quadratically (and not linearly) on $D$.
This implies that 
the leading order effect of the two-body interactions for aligned 
dipolar gases under spherically symmetric harmonic confinement 
is repulsive at high temperature.
Moreover, because $\Delta b_2$ depends quadratically on $D$,
the effects of the dipole-dipole 
interactions will be hard to extract from thermodynamic observables.
It is well known that the generalized
scattering lengths, which characterize the scattering properties
of the two dipole system, can be modified through the application of an 
external magnetic or electric 
field~\cite{YiYou2000,MarinescuYou1998,giovanazzi}.
For fermions,
we show that by reducing one of the generalized scattering lengths,
$\Delta b_2$ depends linearly (and not quadratically) on $D$.
This implies that
the effects of the dipole-dipole interactions can be enhanced
by tuning one of the generalized scattering lengths to near zero.

The remainder of this paper is organized as follows.
Section~\ref{sec_system}
introduces the virial equation of state for a single-component quantum
gas,
the two-body Hamiltonian for aligned dipoles, 
and the different models used
to treat the interaction between the dipoles.
Section~\ref{sec_dipscatt}
studies the low-energy dipole-dipole scattering properties
and characterizes the behavior of the generalized scattering lengths.
Special attention is paid to low-energy shape resonances.
Our main results are presented in Secs.~\ref{sec_results} and \ref{sec_thermo}.
Section~\ref{sec_results} determines the virial coefficient
$\Delta b_2$ for dipolar Bose and Fermi gases under spherically symmetric
confinement within various approximations.
The resulting virial coefficients are used in Sec.~\ref{sec_thermo}
to determine thermodynamic observables such as the total energy
and free energy.
Finally, Sec.~\ref{sec_conclusion} concludes.

\section{Virial equation of state and system under study}
\label{sec_system}
For a single component gas, where the chemical potential $\mu$
is held fixed, the thermodynamic potential $\Omega$ 
in the grand canonical ensemble reads~\cite{DMcQuarrie,KHuang}
\begin{align}
\label{eq_grandpotential}
\Omega = -\beta^{-1} \ln \mbox{Tr} \; e^{-\left(H_n-\mu n\right)\beta},
\end{align}
where $H_n$ is the Hamiltonian of the $n$-body system,
$\beta$ is $(k_BT)^{-1}$, $k_B$ is Boltzmann's constant, $T$ 
is the temperature, and Tr is the trace operator.
In the high-temperature limit,
the chemical potential becomes large and negative,
such that the fugacity $z$, $z=\exp(\mu\beta)$,
is a small parameter.
Taylor expanding Eq.~\eqref{eq_grandpotential} in the 
high-temperature limit, i.e., for $z \ll 1$, 
yields the 
thermodynamic potential in terms of the virial coefficients $b_n$,
\begin{align}
\label{eq_veos}
\Omega \approx -\beta^{-1} Q_1 \sum_{n=1}^{\infty} b_n z^n. 
\end{align}
The $b_n$'s are determined by 
the canonical partition functions $Q_n$, 
\begin{align}
\label{eq_Qn}
Q_n = \mbox{Tr} \; e^{-H_n\beta}.
\end{align}
We evaluate the trace by inserting a
complete set of eigenstates 
with eigenenergies $E_n^{(j)}$, yielding
\begin{align}
\label{eq_QnE}
Q_n = \sum_j e^{-E_n^{(j)}\beta},
\end{align}
where the summation index $j$ collectively denotes the complete
set of quantum numbers allowed by symmetry.

To isolate the effect of the two-body interactions,
we express the grand canonical potential 
$\Omega$ in terms of 
the grand canonical potential $\Omega^{\rm NI}$ 
of the noninteracting system 
plus corrections $\Delta\Omega$~\cite{LiuPRL2009,daily2012a},
\begin{align}
\Omega = \Omega^{\rm NI} + \Delta\Omega.
\end{align}
We are interested in accounting for the leading-order effect of the 
two-body interactions. 
Correspondingly, we truncate Eq.~\eqref{eq_veos} at $n=2$.
Collecting terms of the same order in the fugacity,
the second-order virial coefficient reads
\begin{align}
b_2 = b_2^{\rm NI} + \Delta b_2,
\end{align}
where
$b_2^{\rm NI}$ characterizes the noninteracting
two-body system and $\Delta b_2$,
\begin{align}
\label{eq_db2}
\Delta b_2 = \frac{Q_2 - Q_2^{\rm NI}}{Q_1},
\end{align}
contains the leading-order effect of the two-body interactions.
In Eq.~(\ref{eq_db2}), $Q_2$ and $Q_2^{\rm NI}$ denote the canonical partition
functions of the interacting and noninteracting two-body systems,
respectively.
For spherically symmetric harmonic confinement
with angular trapping frequency $\omega$, as considered throughout this
paper, we find~\cite{bn_explanation}
\begin{align}
\label{eq_bn_ni}
b_n^{\rm NI} = \frac{(\pm 1)^{n+1}}{n}
\frac{Q_1(n\beta)}{Q_1(\beta)},
\end{align}
where
\begin{align}
\label{eq_q1}
Q_1(\beta) = 
\left(\frac{e^{E_{\rm ho}\beta/2}}{e^{E_{\rm ho}\beta}-1}\right)^3.
\end{align}
In Eq.~\eqref{eq_bn_ni}, 
the plus sign is for identical bosons
while the minus sign is for identical fermions.
We define the dimensionless quantity
$\tilde{\omega}$ as the ratio of the harmonic oscillator
energy $E_{\rm ho}$,
$E_{\rm{ho}}=\hbar \omega$ 
(i.e., the characteristic energy of the noninteracting
system), and the thermal energy scale $\beta^{-1}$,
$\tilde{\omega}=E_{\rm ho}\beta$.

To determine the second-order virial coefficient
$\Delta b_2$ that accounts for the interactions between the dipoles
[see Eq.~(\ref{eq_db2})],
we need to determine $Q_2$, i.e., we need 
to determine the eigenenergies $E_2^{(j)}$ of the interacting two-body 
system [see Eq.~(\ref{eq_QnE})].  
Since the two-body interaction depends only on the distance
vector ${\bf{r}}$, the center of mass and relative motion separate,
implying $E_2=E_2^{\text{cm}}+E_2^{\text{rel}}$.
The center of mass contribution has been determined in the 
literature~\cite{LiuPRL2009,daily2012a}
and we thus focus in the following on the relative contribution.
The Schr\"odinger equation for the relative distance vector
${\bf{r}}$ 
reads
\begin{align}
\label{eq_SE}
\left[
-\frac{\hbar^2}{2\mu_{\rm red}}\nabla_{{\bf{r}}}^2
+ \frac{1}{2}\mu_{\rm red}\omega^2 r^2 + V_{\rm int}({\bf{r}})
\right]
\Psi({\bf r})
= E_2^{\rm rel}\Psi({\bf r}),
\end{align}
where $\mu_{\rm red}$ denotes the reduced mass and $V_{\rm int}({\bf{r}})$ the
two-body interaction potential (see below for details).
In what follows,
we assume that the dipoles are aligned by an external field.
This implies that $V_{\rm int}$ is azimuthally symmetric
and that the projection quantum number $m$ is a good quantum number.
The relative orbital angular momentum
$l$, in contrast, is not a good quantum number.
Specifically, 
the interaction potential couples angular momentum 
channels with the same parity.
For identical bosons, the even relative angular momenta $l$ are coupled.
For identical fermions, in contrast, the odd relative
angular momenta $l$ are coupled.

We solve Eq.~\eqref{eq_SE} using two different representations
of the interaction potential.
In the first approach, 
$V_{\rm int}$ is given by~\cite{Bortolotti2006}
\begin{align}
\label{eq_vmodel}
V_{\rm model}({\bf r}) = 
\left\{
\begin{array}{ll}
d^2\frac{1-3\cos^2\theta}{r^3} & \mbox{if $r \ge r_c$} \\
\infty & \mbox{if $r < r_c$} 
\end{array}
\right.,
\end{align}
where $\theta$ denotes the angle between the distance vector
${\bf{r}}$ and the axis along which the dipoles are aligned.
The model potential $V_{\rm model}$ is characterized by two length scales, 
the hard wall radius (or short-range length) $r_c$ and the dipole length
(or long-range length) $D$, $D=\mu_{\rm red} d^2 / \hbar^2$,
where $d$ denotes the dipole moment.
In the second approach, 
we use the regularized zero-range pseudopotential 
$V_{\rm pp,reg}^m$~\cite{deri2003,deri2005,kanj2007},
\begin{align}
\label{eq_pp_full}
V_{\rm pp,reg}^m({\bf r}) = \sum_{l,l'}V_{l,l'}^m({\bf r}).
\end{align}
Equation~\eqref{eq_pp_full} contains an infinite number of terms.
The $V_{l,l'}^m({\bf r})$ are defined by how they act on a wavefunction, 
\begin{align}
\label{eq_pp_part}
V_{l,l'}^m({\bf r}) & \Phi({\bf r}) = \frac{\hbar^2}{2\mu_{\rm red}}
\frac{a_{l,l'}^m}{k^{l+l'}}
\frac{(2l+1)!!(2l'+1)!!}{(2l+1)!}
\frac
{\delta(r)}
{r^{l'+2}} \times \nonumber \\
& Y_{l'm}(\hat{r})
\left[ \frac{\partial^{2l+1}}{\partial r^{2l+1}}r^{l+1}
\int Y_{lm}^*(\hat{r})\Phi({\bf r}) d\hat{r} \right]_{r\to0}, 
\end{align}
where the $Y_{lm}$ denote the spherical harmonics
and where $\hbar k$ is the relative
scattering momentum.
In the absence of the confinement, 
$k$ is given by the relative scattering energy $E_{\rm scatt}^{\rm rel}$, 
$k^2 \hbar^2= 2 \mu_{\rm{red}} E_{\rm scatt}^{\rm rel}$.
In the presence of the harmonic trap, $k$ is determined,
in a self-consistent manner, by the eigenenergies
$E_2^{\rm rel}$ of Eq.~(\ref{eq_SE})~\cite{kanj2007}.
Since $m$ is a good quantum number, each $m$-channel is treated
by a $m$-specific pseudopotential;
for details on the $m=0$ case, see Ref.~\cite{kanj2007}.
The generalized scattering lengths $a_{l,l'}^m$
depend on the dipole moment $d$.
To use $V_{\rm pp,reg}^m$, the $a_{l,l'}^m$ have to be provided as input.
The next section discusses how these input parameters are
determined.

\section{Two-dipole scattering}
\label{sec_dipscatt}
This section determines the generalized free-space 
scattering lengths $a_{l,l'}^m$ for
the shape-dependent model potential $V_{\rm model}$, Eq.~\eqref{eq_vmodel}.
To determine the $a_{l,l'}^m$, 
we consider the relative free-space Schr\"odinger equation
[Eq.~\eqref{eq_SE} with $\omega=0$]
and propagate the logarithmic derivative matrix using
Johnson's algorithm~\cite{john1973} 
with an adaptive step size.
At large $r$,
the logarithmic derivative matrix is matched to the asymptotic
solution and the generalized scattering lengths
are determined from the partial wave phase shifts
$\delta_{l,l'}^m$,
$a_{l,l'}^m(k)=-\tan[\delta_{l,l'}^m(k)]/k$~\cite{yi63}.

Before presenting our numerical results, 
we briefly review the expected low-energy properties of the $a_{l,l'}^m$.
Away from resonances, 
the generalized scattering lengths $a_{l,l'}^m$ [$(l,l') \ne (0,0)$]
are, in general, well approximated by applying the 
first-order Born approximation to the dipole-dipole potential
$V_{\rm dd}$, $V_{\rm dd}=d^2(1-3\cos ^2 \theta)/r^3$~\cite{yi63}.
The Born approximation yields
\begin{align}
\label{eq_all}
a_{l,l}^m = \frac{-2 D [l(l+1)-3m^2]}{(2l-1)(2l+3)l(l+1)}
\end{align}
and 
\begin{align}
\label{eq_allp}
a_{l,l-2}^m = \frac{-D \sqrt{(l^2-m^2)[(l-1)^2-m^2]}}
{(2l-1)l(l-1)\sqrt{(2l+1)(2l-3)}}.
\end{align}
All other $a_{l,l'}^m$ with $(l,l') \ne (0,0)$ are zero.
For $m=0$, Eqs.~\eqref{eq_all} and~\eqref{eq_allp} agree with
Eqs.~(9) and (10) of Ref.~\cite{kanj2007}.
It is important to note that the scattering length
$a_{0,0}^0$ 
cannot, in general, be determined reliably within the
first-order Born approximation~\cite{Bortolotti2006}.

In the following, 
we focus on the scattering properties
of two identical fermionic dipoles, 
implying that $l$ and $l'$ are odd, with $m=0$.
\begin{figure}
\vspace*{+1.5cm}
\includegraphics[angle=0,width=70mm]{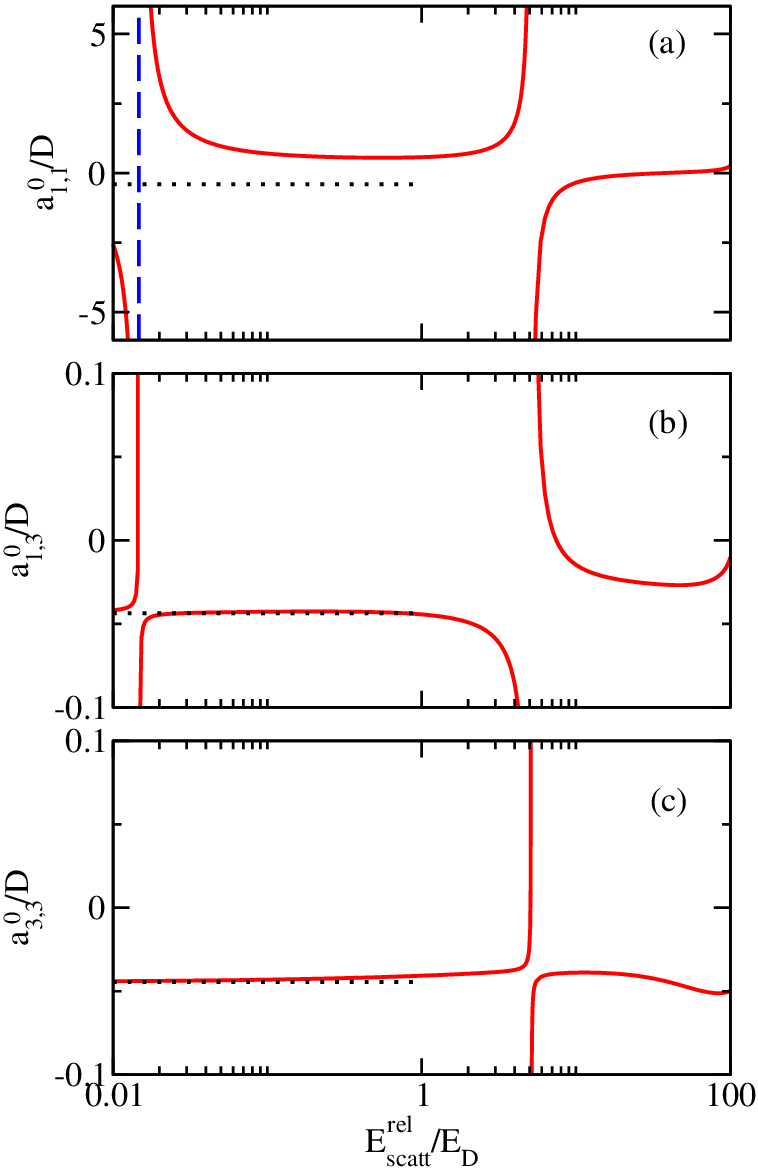}
\vspace*{0.1cm}
\caption{(Color online) 
Generalized scattering lengths (a) $a_{1,1}^0$,
(b) $a_{1,3}^0$, and (c) $a_{3,3}^0$ as a function of 
the relative scattering energy $E_{\rm scatt}^{\rm rel}$
for $D=10.5r_c$.
The dotted lines indicate the zero-energy Born
approximation values of $a_{l,l'}^m$.
In (a), the vertical dashed line indicates
the location of the shape resonance.
Note the different vertical scales of panel~(a) and
of panels~(b) and (c).
}\label{fig_scl}
\end{figure}
Solid lines in Fig.~\ref{fig_scl} 
show the numerically obtained generalized scattering lengths
$a_{l,l'}^0$ with $l',l\leq3$ for $V_{\rm model}$ with $D=10.5r_c$
as a function of the scattering energy $E_{\rm scatt}^{\rm rel}$.
The scattering energy has been scaled by the dipole energy $E_D$,
$E_D=\hbar^2/(\mu_{\rm red}D^2)$.
For comparison, 
dotted horizontal lines show the generalized scattering lengths
predicted by the Born approximation.
The numerically obtained scattering lengths $a_{1,3}^0$ and 
$a_{3,3}^0$ [see Figs.~\ref{fig_scl}(b) and \ref{fig_scl}(c)]
are well approximated by the Born approximation estimates
over a wide range of scattering energies. 
Deviations occur in narrow energy windows around
$E_{\rm scatt}^{\rm rel}/E_D \approx 0.015$ and $5$.
The deviations between the exact values
and the corresponding Born approximation 
values are larger for $a_{1,1}^0$ [see Fig.~\ref{fig_scl}(a)]
than for $a_{3,1}^0$ and $a_{3,3}^0$.
The deviations from the Born approximation values
can be attributed to a quasi-bound state that is, 
approximately, tied to the $l=1$ channel
[see dashed line in Fig.~\ref{fig_scl}(a)].

\begin{figure}
\vspace*{+1.5cm}
\includegraphics[angle=0,width=70mm]{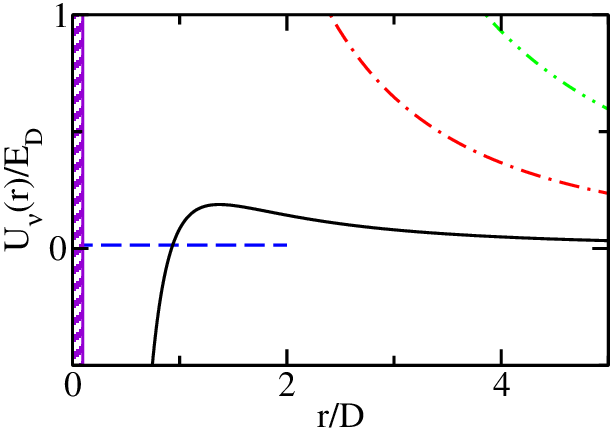}
\vspace*{0.1cm}
\caption{(Color online) 
Adiabatic potential curves
$U_{\nu}(r)$
for two aligned fermionic dipoles.
Though the angular momentum $l$ is not a good quantum number,
the curves can be approximately labeled by $l$.
Solid, dash-dotted, and dash-dot-dotted lines are for
$l \approx 1,3$, and $5$, respectively.
The hard wall is indicated 
by a hatched region extending from zero to 
$r_c=D/10.5$. 
The horizontal dashed line indicates the energy of the 
quasi-bound state.
}\label{fig_potcurves}
\end{figure}

To understand the emergence of the quasi-bound state,
we treat Eq.~\eqref{eq_SE} using $V_{\rm model}$ 
in the adiabatic representation~\cite{ticknor2005,Kanjilal2008}.
Figure~\ref{fig_potcurves}
shows the lowest three adiabatic potential curves 
for two aligned fermionic dipoles with $r_c=D/10.5$
as a function of the distance coordinate $r$.
Though the angular momentum $l$ is not a good quantum number,
each curve can be labeled approximately by $l$.
Solid, dash-dotted, and dash-dot-dotted lines are for $l\approx 1,3,$
and $5$, respectively.
The $l \approx 1$ potential curve exhibits a 
barrier of height around $0.2 E_{D}$,
indicating that the true bound state---which is to
a good approximation described by the lowest potential 
curve~\cite{ticknor2005,Kanjilal2008}---turns into a 
quasi-bound state 
with positive energy after crossing the zero-energy threshold.

\begin{figure}
\vspace*{+1.5cm}
\includegraphics[angle=0,width=70mm]{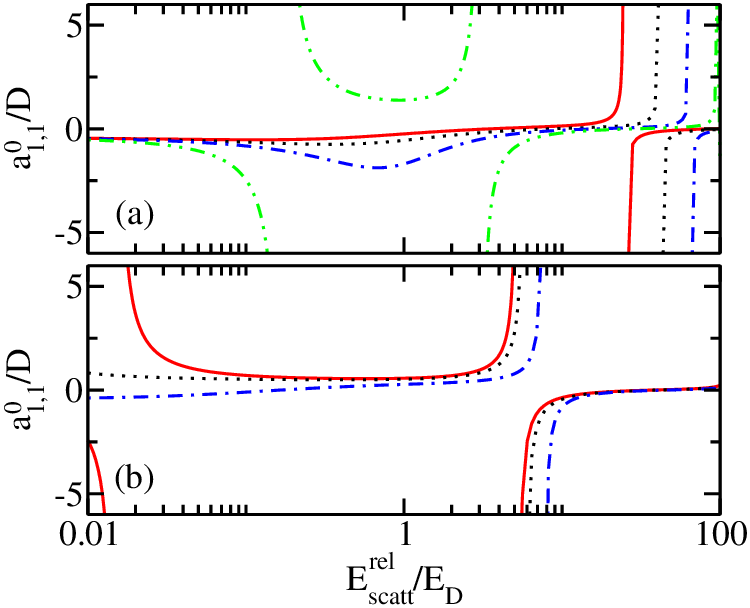}
\vspace*{0.1cm}
\caption{(Color online) 
$a_{1,1}^0$ as a function of the relative 
scattering energy $E_{\rm scatt}^{\rm rel}$ for various values of $D/r_c$.
(a) Solid, dotted, dash-dotted, and dash-dot-dotted lines are for 
$D/r_c=7,8,9,$ and $10$, respectively.
(b) Solid, dotted, and dash-dotted lines are for 
$D/r_c=10.5$ [see also Fig.~\ref{fig_scl}(a)], $10.55,$ and $11$, 
respectively.
}\label{fig_a110}
\end{figure}
Figure~\ref{fig_a110} shows 
the energy dependence of $a_{1,1}^0$ 
for various ratios between the long-range and short-range length
scales in the vicinity of a shape resonance.
For $D/r_c$ values, 
for which no quasi- or shallowly-bound states exist, 
the Born approximation is valid for $E \lesssim E_D$ 
[see the solid and dotted lines in Fig.~\ref{fig_a110}(a)].
As $D/r_c$ increases,
$a_{1,1}^0$ develops a ``dip'' around $E^{\text{rel}} \approx 0.7 E_D$
[see the dash-dotted line in Fig.~\ref{fig_a110}(a)].
At yet larger $D/r_c$, 
the dip turns into two resonance-like features, i.e.,
$a_{1,1}^0$ diverges at energies above and below
the barrier of the $l \approx 1$ adiabatic potential curve 
[see the dash-dot-dotted and solid lines 
in Figs.~\ref{fig_a110}(a) and \ref{fig_a110}(b), respectively].
We associate the resonance features with the appearance of 
a quasi-bound state that ``lives''
to the left of the barrier of the adiabatic potential curve
(see Fig.~\ref{fig_potcurves}).
As $D/r_c$ increases further, 
the energy of the quasi-bound state
moves to smaller (positive) energies
[see the solid line in Fig.~\ref{fig_a110}(b)].
For $D/r_c=10.55$, 
$a_{1,1}^0$ is nearly constant 
for $0.01E_D \lesssim E_{\text{scatt}}^{\text{rel}} \lesssim E_D$
[see the dotted line in Fig.~\ref{fig_a110}(b)].
For larger $D/r_c$,
the quasi-bound state turns into a true bound state and
$a_{1,1}^0$ ``returns'', in the low energy regime,
to its Born approximation value
[see dash-dotted line in Fig.~\ref{fig_a110}(b)].

In the next section, we make two assumptions.
(i)
We assume that one of the generalized scattering lengths is tunable 
while the other generalized scattering lengths are well described
by their Born approximation values.
(ii)
We assume that all generalized scattering lengths are, to 
a good approximation, independent of energy.
To fulfill these assumptions approximately, the external field
should be tuned to a value near resonance. 
Experiments on fermionic alkali 
atoms~\cite{ticknor2004,guenter2005} have
demonstrated that the regime where only one
of the scattering lengths differs notably from its Born approximation
value can be reached.
The requirement to sit near but not on resonance allows one to take
advantage of the fact that the resonance feature of the tunable
scattering length is much broader than that of the other scattering
lengths; this implies that the energy-dependence of the
scattering lengths approximated by their Born approximation values
is weak. Lastly, as shown by the
dotted line in Fig.~\ref{fig_a110}(b) for $a_{1,1}^0$, 
the near-resonance regime also allows one to realize a regime where the
tunable scattering length is roughly independent of energy 
over a wide energy regime.

\section{Second-order virial coefficient $\Delta b_2$}
\label{sec_results}
This section discusses the virial coefficient $\Delta b_2$ 
that accounts for the two-body interactions
for two identical bosons and two identical
fermions as a function of the temperature.
To determine $\Delta b_2$, we take advantage of the fact that the 
relative eigenenergies $E_2^{\rm rel}$ for the trapped two-dipole system 
interacting through $V_{\rm pp,reg}^m$ can, 
provided the generalized scattering lengths
$a_{l,l'}^m$ are known, 
be obtained by solving a transcendental equation~\cite{kanj2007}.
As indicated in our discussion
at the end of Sec.~\ref{sec_dipscatt}, 
we solve the transcendental equation assuming that the
generalized scattering lengths are energy-independent.
We have checked that the use of an energy-dependent
pseudo-potential leaves our main conclusions unaltered.
Combining $E_2^{\rm rel}$ and $E_2^{\text{cm}}$, 
the canonical partition function $Q_2$ [see Eq.~\eqref{eq_QnE}]
and correspondingly the virial coefficient $\Delta b_2$
can be calculated. 
Since the relative two-body energy spectrum is,
in practice,
only known up to a maximum energy $E_2^{\rm rel,max}$,
$\Delta b_2$ is only accurate up to a maximum temperature
or down to a minimum dimensionless inverse temperature $\tilde{\omega}$.
To distinguish the second-order virial coefficient for two 
identical bosons and two identical fermions, 
we use the superscripts $\rm{(B)}$ and $\rm{(F)}$, respectively.

\begin{figure}
\centering
\vspace*{+1.5cm}
\includegraphics[angle=0,width=70mm]{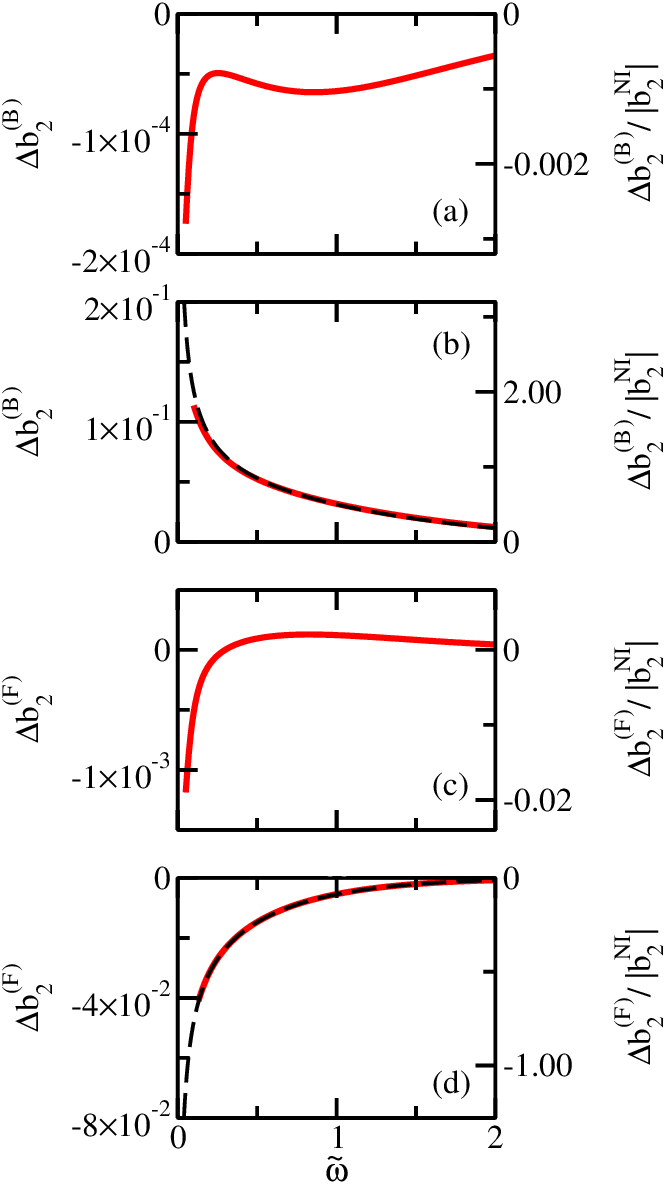}
\vspace*{0.1cm}
\caption{(Color online) Solid lines show $\Delta b_2$ 
as a function of $\tilde{\omega}$ for $D=0.1a_{\rm ho}$ for
(a) and (b) identical dipolar bosons, and 
(c) and (d) identical dipolar fermions.
For comparison, the right vertical axis is scaled by the magnitude
of the noninteracting virial coefficient at $\tilde{\omega}=0$.
(a) All generalized scattering lengths for $l>0$ are given by their
Born approximation values and $a_{0,0}^{0}=0$.
(b) All generalized scattering lengths for $l>0$ are given by their
Born approximation values and $a_{0,0}^{0}=-0.1a_{\rm ho}$; 
the dashed line shows Eq.~\eqref{eq_b2b}.
(c) All generalized scattering lengths are given by their
Born approximation values.
(d) All generalized scattering lengths are given by their
Born approximation values except $a_{1,1}^{0}=0$; 
the dashed line shows Eq.~\eqref{eq_b2f}.
}\label{fig_b20all}
\end{figure}

We start our discussion by considering two identical bosons.
The solid lines in Figs.~\ref{fig_b20all}(a) and \ref{fig_b20all}(b)
show the virial coefficient $\Delta b_2^{\rm (B)}$ for 
a relatively small dipole length $D$, $D=0.1a_{\rm ho}$
[$a_{\rm ho}$ denotes the oscillator length, 
$a_{\rm ho}=\sqrt{\hbar/(\mu_{\rm red}\omega)}$].
To determine the relative eigenenergies $E_2^{\rm rel}$,
the generalized scattering lengths $a_{l,l'}^m$ with $(l,l') \ne (0,0)$
are approximated by the Born approximation values
[see Eqs.~(\ref{eq_all}) and (\ref{eq_allp})].
As discussed in Sec.~\ref{sec_dipscatt},
the $s$-wave scattering length $a_{0,0}^0$ cannot be determined
within the Born approximation. 
Figure~\ref{fig_b20all}(a) assumes that $a_{0,0}^0=0$. 
Correspondingly, 
the resulting virial coefficient is that for a ``purely'' dipolar system.
Figure~\ref{fig_b20all}(b) assumes that $a_{0,0}^0=-0.1a_{\rm ho}$.
Thus, 
the virial coefficient shown in Fig.~\ref{fig_b20all}(b)
reflects the combined effect of the long-range dipole-dipole
interaction and the short-range $s$-wave interaction.
For the $\tilde{\omega}$ values shown,
the solid lines are converged to better than 1\% and include all states
with relative energy less than $51.5\hbar\omega$,
that is, $E_2^{\rm rel,max} < 51.5\hbar\omega$.

For the Born approximation to be applicable, 
two conditions have to be met.
First, 
the virial coefficient $\Delta b_2^{\rm (B)}$
has to be converged for an energy cutoff $E_2^{\rm rel,max}$ that is smaller
than $E_D$. 
This condition comes from the
fact that the Born approximation is only valid for
$E_{\rm scatt}^{\rm rel} \lesssim E_D$.
In Figs.~\ref{fig_b20all}(a) and \ref{fig_b20all}(b), 
we have $E_D \approx 100 E_{\rm ho}$ and thus
$E_2^{\rm rel,max}$ and $E_{\rm scatt}^{\rm rel} < E_D$.
Second, 
the system has to be away from shape resonances that are associated 
with quasi-bound states whose energy lies below $E_{2}^{\rm rel,max}$.

The second-order virial coefficient $\Delta b_2^{\rm (B)}$ for the purely
dipolar Bose gas under spherically symmetric harmonic
confinement [Fig.~\ref{fig_b20all}(a)]
is negative for all $\tilde{\omega}$ shown,
indicating that the leading-order effect of the dipole-dipole 
interaction is repulsive for aligned Bose gases with $D/a_{\rm ho}$ 
small compared to one.
Figure~\ref{fig_b20all}(a) shows that
the effect of the two-body interaction is very small.
Specifically, 
the contribution $\Delta b_2^{\rm (B)}$ is about three orders of magnitude
smaller than the noninteracting contribution
$b_2^{\rm NI}$ at $\tilde{\omega}=0$
[see the right axis of Fig.~\ref{fig_b20all}(a)], 
suggesting that the effects of the dipole-dipole interactions would 
be hard to extract from thermodynamic observables
for a purely dipolar gas with small $D/a_{\rm ho}$.
The virial coefficient for $a_{0,0}^0=-0.1a_{\rm ho}$ 
[Fig.~\ref{fig_b20all}(b)]
is about three orders of magnitude larger than that for 
$a_{0,0}^0=0$ [Fig.~\ref{fig_b20all}(a)]. 
Moreover, 
the sign is reversed.
Since the dipole length $D$ is the same in 
Figs.~\ref{fig_b20all}(a) and \ref{fig_b20all}(b),
it is clear that the increase of $\Delta b_2^{\rm (B)}$ 
is due to the change in $a_{0,0}^0$. 
The fact that $\Delta b_2^{\rm (B)}$ is positive 
for $a_{0,0}^0=-0.1a_{\rm ho}$ makes 
sense intuitively since a positive $\Delta b_2^{\rm (B)}$ 
implies an effective attraction as would be naively 
expected for a negative $a_{0,0}^0$.

We now consider two identical fermions.
The solid lines in Figs.~\ref{fig_b20all}(c) 
and \ref{fig_b20all}(d) 
show $\Delta b_2^{\rm (F)}$ for two identical dipolar fermions
for $D=0.1a_{\rm ho}$.
In Fig.~\ref{fig_b20all}(c), 
the energies $E_2^{\rm rel}$ are obtained by approximating all
generalized scattering lengths by the Born approximation values.
Thus, 
the virial coefficient shown in Fig.~\ref{fig_b20all}(c) is that
for a purely dipolar gas under spherically symmetric external confinement
with small $D/a_{\rm ho}$.
$\Delta b_2^{\rm (F)}$ is negative at high temperature,
changes sign at $k_B T \approx E_{\rm ho}/0.2$, 
and is positive for smaller temperatures (larger $\tilde{\omega}$).
In Fig.~\ref{fig_b20all}(d), 
in contrast,
the energies $E_2^{\rm rel}$ are obtained by approximating all
but one of the generalized scattering lengths by their
Born approximation value.
Specifically,
we choose $a_{1,1}^{0}=0$, i.e., 
we assume that this generalized scattering length is tuned away from its 
Born approximation value through the application of an external field.
Figure~\ref{fig_b20all}(d) shows that $\Delta b_2^{\rm (F)}$ 
is negative for all temperatures and 
that $|\Delta b_2^{(\text{F})}|$ is between one and two
orders of magnitude larger than 
the absolute value of the virial coefficient for
the purely dipolar gas [see Fig.~\ref{fig_b20all}(c)].
Since the dipole length $D$ is the same in 
Figs.~\ref{fig_b20all}(c) and \ref{fig_b20all}(d),
it is clear that the increase of $|\Delta b_2^{\rm (F)}|$ 
is due to the change in $a_{1,1}^0$. 
Figure~\ref{fig_scl} shows that the scenario
considered here can be realized, 
at least approximately, 
for the model potential $V_{\rm model}$.
For the parameters considered in Fig.~\ref{fig_scl},
the $a_{l,l'}^m$ with $(l,l') \ne (1,1)$ are, 
except for a tiny energy window, 
well approximated by the Born approximation value for 
$E_{\text{scatt}}^{\text{rel}}$ up to 
$E_{\text{scatt}}^{\text{rel}} \approx E_D$.
Moreover, the discussion around Fig.~\ref{fig_a110} shows that
$a_{1,1}^0$ can, at least approximately, be tuned to a value
different from the Born approximation value while all other
$a_{l,l'}^m$ are quite well described by their Born approximation value.

We now develop a perturbative approach that 
(i) shows analytically that the second-order virial coefficients 
$\Delta b_2^{\rm (B)}$ and $\Delta b_2^{\rm (F)}$ 
for the purely dipolar gases do not contain
terms linear in $D$, 
thereby explaining the small values of 
$\Delta b_2^{\rm (B)}$ and $\Delta b_2^{\rm (F)}$ 
in Figs.~\ref{fig_b20all}(a) and \ref{fig_b20all}(c),
and 
(ii) yields analytic expressions in the case where one or more of
the generalized scattering lengths are assumed to be tunable 
[see dashed lines in Figs.~\ref{fig_b20all}(b) and \ref{fig_b20all}(d)].
The idea is to calculate the relative energies 
$E_2^{\rm rel}$ perturbatively and
to then use the perturbative energy expressions to calculate
$\Delta b_2$ analytically to linear order in $D/a_{\rm ho}$.

We assume that $D/a_{\rm ho}$ is small
and employ first-order degenerate perturbation theory.
Specifically, 
we treat the noninteracting harmonic oscillator Hamiltonian 
in the relative degrees of freedom as the unperturbed Hamiltonian and
the regularized pseudopotential $V_{\rm pp,reg}^m$ as a perturbation.
We label the unperturbed states $\psi_{n,l,m}^{(0)}$ 
by the principal quantum number $n$, 
the orbital angular momentum quantum number $l$ 
and the projection quantum number $m$,
and find compact analytical expressions for the
matrix elements
$\langle V_{nl,n'l'}^m \rangle = 
\langle \psi_{n,l_1,m_1}^{(0)} | V_{l,l'}^m | \psi_{n',l_2,m_2}^{(0)} \rangle$,
\begin{align}
\label{eq_pert_matrix_element}
\langle V_{nl,n'l'}^m \rangle 
= & \delta_{m,m_1} \delta_{m,m_2} \delta_{l',l_1} \delta_{l,l_2} 
\frac{2^{2+l+l'}}{\pi} \frac{a_{l,l'}^m}{k^{l+l'}a_{\rm ho}^{l+l'+1}} 
\\ \nonumber & \times
\sqrt{\frac{\Gamma(n+l+3/2)}{n!}} \sqrt{\frac{\Gamma(n'+l'+3/2)}{n'!}}
\; E_{\rm ho}. 
\end{align}
Diagonalizing the perturbation matrix spanned by the matrix elements
$\langle V_{nl,n'l'}^m \rangle $ 
for each noninteracting energy manifold, i.e.,
for the manifolds with unperturbed energies
$3 E_{\rm ho}/2$, $5 E_{\rm ho}/2$, and so on,
we find the first-order energy shifts
$\Delta E_{n,l,m}^{(1)}$, $\Delta E_{n,l,m}^{(1)}=c_{n,l,m} E_{\rm ho}$.
Using these energy shifts in
Eqs.~\eqref{eq_Qn} and~\eqref{eq_db2},
the second-order virial coefficient in the weakly-interacting
regime can be written as
\begin{align}
\label{eq_db2_perturbative}
\Delta b_2 \approx -\tilde{\omega} 
\sum_{nlm} c_{nlm} e^{-(2n+l+3/2)\tilde{\omega}}.
\end{align}
So far, the discussion is general, i.e., we have not yet 
specified whether we are dealing with bosons or fermions.
Correspondingly, Eq.~(\ref{eq_db2_perturbative}) writes $\Delta b_2$
without the $\rm{(B)}$/$\rm{(F)}$
superscript.
Inspection of Eq.~(\ref{eq_pert_matrix_element}) shows that the matrix
elements $\langle V_{nl,n'l'}^m \rangle$ depend on $m$ 
only through the generalized scattering lengths.
Using this and rewriting the sum of eigenvalues for each energy manifold
in terms of the trace of the corresponding perturbation matrix, 
we find
\begin{align}
\label{eq_db2_perturbative2}
\Delta b_2 \approx & -\tilde{\omega} 
\sum_{nl} e^{-(2n+l+3/2)\tilde{\omega}} 
 \nonumber\\ & \times
\frac{2^{l+2} \Gamma(n+l+3/2)}{(2n+l+3/2)^l \pi n!}
\left( \sum_{m=-l}^l \frac{a_{l,l}^m}{a_{\rm ho}} \right),
\end{align}
where we have replaced the scattering momentum $\hbar k$ 
by the unperturbed energies, 
that is, we used $k^2=2(2n+l+3/2)/a_{\rm ho}^2$.
In Eq.~(\ref{eq_db2_perturbative2}),
the sum over $l$ is restricted to even or odd $l$
for bosons or fermions, respectively.

We now apply Eq.~(\ref{eq_db2_perturbative2}) to the purely dipolar
Bose and Fermi gases.
Using the Born approximation values [see Eq.~(\ref{eq_all})],
we find 
\begin{eqnarray}
\label{eq_average}
\sum_{m=-l}^l a_{l,l}^m=0.
\end{eqnarray}
This shows that $\Delta b_2^{(B)}$ and $\Delta b_2^{(F)}$ do not, 
in the small $D/a_{\rm ho}$ regime, 
contain terms that are linear in the dipole length $D$, 
in agreement with the fact that the virial coefficients shown 
in Figs.~\ref{fig_b20all}(a) and \ref{fig_b20all}(c) are very small.
To determine the leading-order dependence of $\Delta b_2$ on $D$ 
for purely dipolar gases, 
our treatment would have to be modified in two places. 
First, 
the generalized scattering lengths would have to be known to
second order in $D$ and, 
second, 
the perturbation treatment would have to be pushed to second order. 
These generalizations are not pursued in this work.

Next, 
we focus on the scenario where one or more of the
generalized scattering lengths are tunable 
[see Figs.~\ref{fig_b20all}(b) and \ref{fig_b20all}(d)].
For identical bosons, 
we assume that the $a_{l,l'}^m$ with $(l,l') \ne (0,0)$
are given by the Born approximation values and that $a_{0,0}^0$ takes a small
but finite value ($|a_{0,0}^0|/a_{\rm ho}$ much smaller than one).
Under these assumptions, 
the perturbative treatment yields
\begin{align}
\label{eq_b2b}
\Delta b_2^{(B)} \approx - \tilde{\omega}\frac{2}{\sqrt{\pi}}
\frac{e^{-3\tilde{\omega}/2}}{(1-e^{-2\tilde{\omega}})^{3/2}}
\frac{a_{0,0}^{0}}{a_{\rm ho}} + 
{\cal O}((D/a_{\rm ho})^2).
\end{align}
Equation~(\ref{eq_b2b}) breaks down in the $\tilde{\omega} \rightarrow 0$
limit but is expected to be valid in the $\tilde{\omega}$ regime considered
in this paper.
The dashed line in Fig.~\ref{fig_b20all}(b) shows
Eq.~(\ref{eq_b2b}).
The agreement with the solid line is excellent, 
indicating that the virial coefficient $\Delta b_2^{(B)}$ is, 
to a very good approximation,
determined by $a_{0,0}^0$ alone and independent of $D$.
This implies that the dipolar Bose gas with tunable $a_{0,0}^0$
behaves, 
to leading order, 
like a Bose gas with isotropic $s$-wave interactions. 
The reason is that the leading order dipolar contributions to
$\Delta b_2^{(B)}$ average to zero 
in the regime where the Born approximation is applicable 
[see Eq.~(\ref{eq_average})].

For identical fermions,
we assume that the scattering lengths with $l=1$ are tunable
and that all other generalized scattering lengths 
are approximated by the Born approximation values.
Using the perturbative energies 
(i.e., assuming $|a_{1,1}^{m}| \ll a_{\rm ho}$)
and evaluating the sums in Eq.~\eqref{eq_db2_perturbative2},
we find 
\begin{align}
\label{eq_b2f}
\Delta b_2^{(F)} \approx & -\tilde{\omega}
\frac{12e^{-5\tilde{\omega}/2}}{5\sqrt{\pi}}
\left(\frac{a_{1,1}^1}{a_{\rm ho}}
+\frac{a_{1,1}^0}{a_{\rm ho}}
+\frac{a_{1,1}^{-1}}{a_{\rm ho}}\right)
\\ \nonumber &\times 
\;_2F_1\!\left(\frac{5}{4},\frac{5}{2},\frac{9}{4},e^{-2\tilde{\omega}}\right)
 + {\cal O}((D/a_{\rm ho})^2),
\end{align}
where $\;_2F_1$ is the hypergeometric function.
Equation~\eqref{eq_b2f} breaks down in the $\tilde{\omega} \to 0$ limit but is
expected to be valid in the $\tilde{\omega}$ regime considered
in this paper. 
The dashed line in Fig.~\ref{fig_b20all}(d) shows Eq.~(\ref{eq_b2f})
with $a_{1,1}^0=0$ and $a_{1,1}^{\pm 1}$ given by their
Born approximation value.
The agreement with the solid line is excellent. 
Our analysis shows that the virial coefficient $\Delta b^{(\text{F})}$
depends linearly on $D$ if one
of the $a_{l,l'}^m$ is tuned away from its Born approximation value.
Perturbative expressions analogous 
to Eqs.~\eqref{eq_b2b} and \eqref{eq_b2f}
can be derived if other  
low partial-wave scattering lengths are assumed to be tunable.

\section{Thermodynamics}
\label{sec_thermo}
\begin{figure}
\vspace*{+1.5cm}
\includegraphics[angle=0,width=70mm]{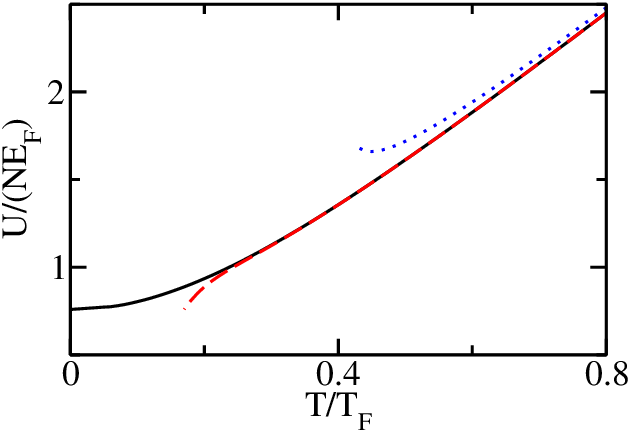}
\vspace*{0.1cm}
\caption{(Color online) 
Total energy per particle $U/N$ for $N=100$ 
identical fermions under spherically symmetric confinement
as a function of the temperature $T$.
The solid line shows $U/N$ for the non-interacting Fermi gas.
Dashed and dotted lines show $U/N$ for the interacting
Fermi gas with $D=0.1 a_{\text{ho}}$
obtained using the virial equation of state 
to second order in the fugacity.
The dashed line shows results for the purely dipolar Fermi gas.
The dotted line shows results for $a_{1,1}^0=0$
while other $a_{l,l'}^m$ are approximated by their
Born-approximation value.
}\label{fig_energy}
\end{figure}

\begin{figure}
\vspace*{+1.5cm}
\includegraphics[angle=0,width=70mm]{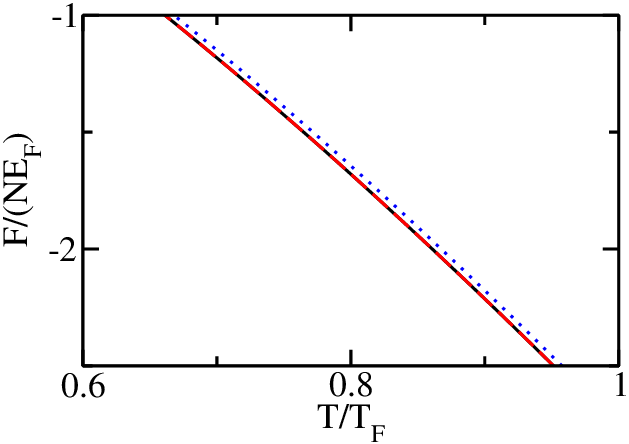}
\vspace*{0.1cm}
\caption{(Color online) 
Helmholtz free energy per particle $F/N$ 
for $N=100$ identical fermions under external spherically symmetric
confinement as a function of the temperature $T$.
The parameters and line styles are the same as in 
Fig.~\protect\ref{fig_energy}.
The dashed and solid lines are indistinguishable on the scale shown.
}\label{fig_freeEnergy}
\end{figure}

This section examines the thermodynamics of aligned dipolar gases
under spherically symmetric harmonic confinement
using the virial equation of state up to second order in the fugacity.
Figures~\ref{fig_energy} and \ref{fig_freeEnergy}
show the total energy per particle $U/N$
and the Helmholtz free energy per particle $F/N$,
respectively, 
as a function of the temperature $T$ 
for $N=100$ identical fermions with $D=0.1a_{\rm ho}$.
Both quantities are scaled by the 
Fermi energy $E_F$, which---in 
the semiclassical approximation---reads
$E_F=(6N)^{1/3}\hbar\omega$.
The dashed lines show $U/N$ and $F/N$ for the purely dipolar Fermi gas
with the $a_{l,l'}^m$ approximated by their Born approximation value.
The dotted lines, in contrast, are for the case where
$a_{1,1}^0$ is assumed to be tunable (here, $a_{1,1}^0=0$)
and where all other $a_{l,l'}^m$ are approximated by their
Born approximation value.
For comparison, 
the solid lines show $U/N$ and $F/N$
for the non-interacting Fermi gas.
On the scale of Figs.~\ref{fig_energy} and \ref{fig_freeEnergy},
the dashed lines are nearly  indistinguishable from the solid lines; 
for temperatures above $k_BT = 0.55 E_F$, 
which is roughly where we expect that the virial equation of state 
yields reliable results, 
the deviations are less than 0.01\%. 
This implies that the effects of the dipolar interactions 
will be hard to observe 
for weakly-interacting purely dipolar gases.
In essence, 
the dipolar effects get averaged over, 
making the weakly-interacting purely dipolar gas behave approximately like
a non-interacting gas.

If the scattering length $a_{1,1}^0$ is tuned to zero 
(see dotted lines in Figs.~\ref{fig_energy} and \ref{fig_freeEnergy}),
the dependence of the thermodynamic 
observables on the dipolar interactions is enhanced.
The difference between the thermodynamic observables 
for the interacting and non-interacting systems is
of the order of 3\% for $k_B T \approx 0.8 E_F$. 
This suggests that dedicated cold atom experiments 
on erbium or dysprosium
might be able
to detect the effects of the dipolar 
interactions in thermodynamic measurements
at temperatures around the Fermi temperature. 
As discussed above, 
the tunability of one of the scattering lengths is
possible if the system under study exhibits non-overlapping resonances.
In practice, 
a given scattering length may change a fair amount 
in the relevant energy range or overlapping resonances may lead to
deviations in more than one scattering length.
While an exhaustive study of all 
possible non-universal parameter combinations is
beyond the scope of this work, 
our key point, namely, 
the fact that tuning one of the scattering lengths
away from its Born approximation value (and possibly to zero)
can enhance the signatures of the dipole-dipole interactions 
of weakly-interacting gases in thermodynamic observables,
should apply also in more complicated situations.

\section{Conclusion}
\label{sec_conclusion}
This paper considered the high-temperature virial equation of state
of weakly-interacting single-component dipolar quantum gases 
under spherically symmetric harmonic confinement.
For purely bosonic and fermionic dipoles,
we found that the lowest-order correction to the ideal gas is repulsive
in the high-temperature limit.
Compared to the non-interacting gases, 
the modifications of the thermodynamic observables are very small.
Using a perturbative approach,
we showed that the corrections are small since the second
order virial coefficient depends quadratically on the dipole length $D$;
terms linear in $D$ average to zero within the Born approximation.
We argued that the effects of the dipole-dipole interactions 
can be enhanced if one or more generalized scattering lengths can be tuned
away from their Born approximation value.
For the fermionic gas, 
we showed explicitly that the generalized scattering length $a_{1,1}^0$
can be tuned in the vicinity of a low-energy shape resonance,
leading to a second order virial coefficient that depends
linearly on the dipole length. 
Correspondingly, 
the effects of the dipolar interaction are, 
compared to the purely dipolar case, 
greatly enhanced.

Throughout, 
we considered single-component Bose and Fermi gases.
If we considered two-component dipolar gases, 
then the thermodynamic properties would depend 
on the interspecies and intraspecies virial coefficients. 
For a two-component Fermi gas, e.g., 
the intraspecies virial coefficients would be given by 
$\Delta b_2^{(\text{F})}$ 
while the interspecies virial coefficient would be given 
by a combination of the even and odd $l$ virial coefficients, i.e., 
of $\Delta b_2^{(\text{B})}$ and $\Delta b_2^{(\text{F})}$.
While the thermodynamic observables would differ from those
considered in this paper, 
our perturbative analysis of the virial
coefficients carries over unchanged.

The present work presents a first step toward understanding the
thermodynamic properties of dipolar systems. 
While the present work provides some insights into weakly-interacting
dipolar gases at finite temperature, 
many questions remain unanswered.
What happens if the dipolar interactions are of intermediate
strength or strong? 
Do these systems exhibit universal thermodynamic properties?
What are the effects of the third- and higher-body virial coefficients?
What are the effects of anisotropic confinement geometries?

{\em{Acknowledgment:}}
Support by the National
Science Foundation (NSF) through Grant Nos.
PHY-1205443 and PHY-1306905
is gratefully acknowledged.


\begin{thebibliography}{10}
\bibitem{bn_explanation}
It can be shown that Eq.~(\protect\ref{eq_bn_ni}) holds
not only for a spherically symmetric harmonic trap but also for
other confining geometries.
\end{thebibliography}
\end{document}